\documentclass[aip,twocolumn]{revtex4}

\usepackage{amssymb}
\usepackage{amsmath}

\def\vb#1{\mbox{\boldmath$#1$}}
\def\pd#1#2{\frac{\partial #1}{\partial #2}}
\def\fd#1#2{\frac{\delta #1}{\delta #2}}
\def\wh#1{\widehat{#1}}
\def\bdot{\,\vb{\cdot}\,}
\def\btimes{\,\vb{\times}\,}

\def\bhat{\wh{{\sf b}}}
\def\exd{{\sf d}}

\newcommand{\bc}{\begin{center}}
\newcommand{\ec}{\end{center}}
\newcommand{\bt}{\begin{tabbing}}
\newcommand{\et}{\end{tabbing}} 
\newcommand{\be}{\begin{eqnarray*}}
\newcommand{\ee}{\end{eqnarray*}}

\begin{document}

\title{Lagrangian and Hamiltonian constraints for guiding-center Hamiltonian theories}

\author{Natalia Tronko$^{1}$ and Alain J.~Brizard$^{2}$}
\affiliation{$^{1}$Max-Planck-Institut f\"{u}r Plasmaphysik, 85748 Garching, Germany \\ 
$^{2}$Department of Physics, Saint Michael's College, Colchester, VT 05439, USA}

\begin{abstract}
A consistent guiding-center Hamiltonian theory is derived by Lie-transform perturbation method, with terms up to second order in magnetic-field nonuniformity. Consistency is demonstrated by showing that the guiding-center transformation presented here satisfies separate Jacobian and Lagrangian constraints that have not been explored before. A new first-order term appearing in the guiding-center phase-space Lagrangian is identified through a calculation of the guiding-center polarization. It is shown that this new polarization term also yields a simpler expression of the guiding-center toroidal canonical momentum, which satisfies an exact conservation law in axisymmetric magnetic geometries. Lastly, an application of the guiding-center Lagrangian constraint on the guiding-center Hamiltonian yields a natural interpretation for its higher-order corrections.
\end{abstract}

\begin{flushright}
October 18, 2015
\end{flushright}

\maketitle

\section{Introduction} 

The consistent derivation of a Hamiltonian guiding-center theory that includes second-order effects in magnetic-field nonuniformity is an important problem in magnetic fusion plasma physics. While the derivation of the second-order corrections in the guiding-center Hamiltonian equations of motion yield higher-order corrections that may be ignored in practical applications, they can nonetheless be useful in gaining insights into higher-order perturbation theory.

\subsection{Previous works}

Recently, Parra and Calvo \cite{Parra_Calvo_2011} and Burby, Squire, and Qin \cite{Burby_SQ_2013} derived guiding-center theories with second-order corrections in the guiding-center Hamiltonian using different methods. Parra and Calvo \cite{Parra_Calvo_2011} constructed their guiding-center transformation based on a {\it microscopic} view that treats the lowest-order gyroradius $\rho_{\rm g}$ as a zeroth-order (nonperturbative) term that is introduced by a preliminary transformation, which introduces explicit gyroangle dependence in the preliminary phase-space Lagrangian. The subsequent derivation of the guiding-center phase-space Lagrangian proceeds through an asymptotic expansion in powers of a small ordering parameter $\epsilon_{\rm B} \equiv \rho_{\rm g}/L_{\rm B} \ll 1$ defined as the ratio of the gyroradius $\rho_{\rm g}$ (which is considered finite in the microscopic view) to the magnetic nonuniformity length scale $L_{\rm B} \gg \rho_{\rm g}$. Burby, Squire, and Qin \cite{Burby_SQ_2013}, on the other hand, derived the second-order guiding-center Hamiltonian through a computer-based algorithm that bypassed the issue of gyrogauge invariance.

These two theories were compared in Ref.~\cite{PCBSQ_2014} and were found to agree up to a gyroangle-independent gauge term in the guiding-center phase-space Lagrangian. Both works (which assume a vanishing electric field ${\bf E} = 0$) reproduced the first-order results of the pioneering work of Littlejohn \cite{RGL_1979,RGL_1981,RGL_1983}, which made certain simplifying assumptions on the symplectic part of the guiding-center phase-space Lagrangian (see Ref.~\cite{Cary_Brizard_2009} for a review of Hamiltonian guiding-center theory).

\subsection{Present work}

The purpose of the present work is to use the standard Lie-transform perturbation method to derive higher-order guiding-center Hamilton equations of motion with as few assumptions about the guiding-center Hamiltonian and Poisson-bracket structure as possible. The consistency of our guiding-center transformation will be checked through Jacobian, Hamiltonian, and Lagrangian constraints. Only results are presented here and details of the calculations are presented elsewhere \cite{Brizard_Tronko_2015}.

In the present work, we recover standard expressions for the guiding-center polarization \cite{Pfirsch_1984,CRPW_1986,Kaufman_1986,Brizard_2013}. We also show that a consistent treatment of a guiding-center polarization and its role in providing a more transparent guiding-center representation of the toroidal canonical angular momentum, which is an exact constant of motion in axisymmetric magnetic geometry, both require that a new first-order term be kept in the symplectic part of the guiding-center phase-space Lagrangian \cite{Brizard_2013}.  

\subsection{Organization}

The remainder of the paper is organized as follows. In Sec.~\ref{sec:HOgc}, equivalent representations of guiding-center Hamiltonian theory are presented in terms of the guiding-center Hamiltonian \eqref{eq:Hamiltonian_gc} and the guiding-center Poisson bracket \eqref{eq:PB_gc_star}, in which the guiding-center magnetic moment $\mu \equiv J\,\Omega/B$ (expressed in terms of the gyroaction $J$) is uniquely defined and higher-order corrections due to magnetic-field nonuniformity are included in either the guiding-center potential energy $\Psi \equiv J\,\Omega + \cdots$ or the guiding-center symplectic momentum $\vb{\Pi} \equiv p_{\|}\,\bhat + \cdots$. In the Hamiltonian representation $(\vb{\Pi} \equiv p_{\|}\bhat)$, these higher-order corrections appear only in the guiding-center Hamiltonian, while, in the symplectic representation $(\Psi \equiv J\,\Omega)$, they appear only in the guiding-center Poisson bracket. 

In Sec.~\ref{sec:Cgc}, the higher-order guiding-center transformation is given up to second order in magnetic-field nonuniformity, and it is shown to simultaneously satisfy several consistency constraints based on the guiding-center Jacobian, Hamiltonian, and Lagrangian. These constraints leave only the perpendicular components of the first-order symplectic momentum $\vb{\Pi}_{1\bot}$ unspecified. In previous works, from Littlejohn's work \cite{RGL_1979,RGL_1981,RGL_1983} up until recent work \cite{Parra_Calvo_2011,Burby_SQ_2013,MSY_2013}, the choice $\vb{\Pi}_{1\bot} \equiv 0$ was implicitly assumed. In Ref.~\cite{Brizard_2013}, it was shown that a new constraint on the choice for $\vb{\Pi}_{1\bot}$ is imposed if the guiding-center transformation must also yield the standard Pfirsch-Kaufman expression for the guiding-center polarization \cite{Pfirsch_1984,CRPW_1986,Kaufman_1986}. This new choice is shown in Sec.~\ref{sec:gcPol_TCM} to lead to a more transparent guiding-center representation for the toroidal canonical momentum, which is an exact constant of motion in axisymmetric tokamak geometry.

\section{\label{sec:HOgc}Higher-order Guiding-center Hamiltonian Theory} 

In the following perturbation analysis, we use the {\it macroscopic} view (i.e., $L_{\rm B}$ is finite and $\rho_{\rm g} \ll L_{\rm B}$), which introduces a dimensionless ordering parameter $\epsilon$ used in renormalizing the electric charge $e \rightarrow e/\epsilon$ (e.g., $\Omega = eB/mc \rightarrow \epsilon^{-1}\Omega$) \cite{Cary_Brizard_2009}. According to this view, a preliminary phase-space transformation is not required and physical results are recovered by setting $\epsilon = 1$ (while ordering in $]epsilon_{B}$ is simply determined by inspection).

\subsection{Guiding-center Hamiltonian and Poisson-bracket structure}

Guiding-center Hamiltonian dynamics is expressed in terms of a guiding-center Hamiltonian function that depends on the guiding-center position ${\bf X}$, the guiding-center parallel momentum $p_{\|}$, and the guiding-center gyroaction $J \equiv \mu\,B/\Omega$; it is, however, independent of the gyroangle $\theta$ at all orders. Since the guiding-center phase-space coordinates are non-canonical coordinates, a noncanonical guiding-center Poisson bracket, whose components are also gyroangle-independent, is also needed. 

The guiding-center Hamiltonian $H_{\rm gc}$ and the guiding-center symplectic structure defined by the Poincar\'{e}-Cartan one-form $\Gamma_{\rm gc}$ (from which the guiding-center Poisson bracket is constructed) are used to construct the guiding-center phase-space Lagrangian:
\begin{eqnarray}
\Lambda_{\rm gc} & \equiv & \Gamma_{\rm gc} \;-\; H_{\rm gc}\,dt \nonumber \\
 & = & \left( {\sf T}_{\rm gc}^{-1}\Gamma_{0} \;+\; \exd S\right) \;-\; \left( {\sf T}_{\rm gc}^{-1}H_{0}\right)\,dt,
\label{eq:Gamma_H_gc}
\end{eqnarray}
where ${\sf T}_{\rm gc}^{-1}$ denotes the guiding-center (push-forward) Lie-transform operator and $S$ denotes an arbitrary gauge-function. In addition, the lowest-order Hamiltonian and symplectic structure
\begin{equation}
\left. \begin{array}{rcl}
H_{0} & \equiv & p_{\|0}^{2}/2m + J_{0}\Omega({\bf x}) \\
 &  & \\
\Gamma_{0} & \equiv & \left[e\,{\bf A}({\bf x})/c \;+\frac{}{} {\bf p}_{0}({\bf x},p_{\|0},J_{0},\theta_{0})\right]\bdot\exd{\bf x}
\end{array} \right\},
\label{eq:HGamma_0}
\end{equation}
 are expressed in terms of the lowest-order guiding-center (local particle) coordinates 
 \begin{equation}
 z_{0}^{\alpha} \equiv ({\bf x},p_{\|0},J_{0},\theta_{0}), 
 \label{eq:z0_def}
 \end{equation}
 where ${\bf p}_{0} \equiv p_{\|0}\bhat ({\bf x}) + {\bf p}_{\bot 0}(J_{0},\theta_{0},{\bf x})$ denotes the local particle momentum expressed in terms of parallel and perpendicular components defined with respect to the magnetic unit vector  $\bhat({\bf x})$ at the particle position ${\bf x}$. 
 
 The guiding-center Euler-Lagrange equations are obtained from the guiding-center variational principle $\delta\int\Lambda_{\rm gc} = 0$:
 \begin{equation}
 \left(\vb{\omega}_{\rm gc}\right)_{\alpha\beta}\;\frac{d_{\rm gc}Z^{\beta}}{dt} \;=\; \pd{H_{\rm gc}}{Z^{\alpha}},
 \label{eq:EL_gc}
 \end{equation}
where the guiding-center Lagrange two-form $\vb{\omega}_{\rm gc} = \exd\Gamma_{\rm gc}$ has the components $(\vb{\omega}_{\rm gc})_{\alpha\beta} \equiv
\partial_{\alpha}\Gamma_{{\rm gc}\beta} - \partial_{\beta}\Gamma_{{\rm gc}\alpha}$, which form an anti-symmetric matrix. We note that the exact one-form $\exd S$ in 
Eq.~\eqref{eq:Gamma_H_gc} does not change the guiding-center Lagrange two-form $\vb{\omega}_{\rm gc} = \exd\Gamma_{\rm gc} =  {\sf T}_{\rm gc}^{-1}(\exd\Gamma_{0}) = {\sf T}_{\rm gc}^{-1}\vb{\omega}_{0}$, since the exterior derivative $\exd$ satisfies the identity $\exd^{2}S \equiv 0$ (analogous to the vector identity $\nabla\btimes\nabla S = 0$), and $\exd$ commutes with ${\sf T}_{\rm gc}^{-1}$.  

\subsubsection{Equivalent Hamiltonian theories}

In the present work, the guiding-center Hamiltonian in Eq.~\eqref{eq:Gamma_H_gc} is defined as
\begin{equation}
H_{\rm gc} \;\equiv\; \frac{p_{\|}^{2}}{2m} \;+\; \Psi,
\label{eq:Hamiltonian_gc}
\end{equation}
where the effective guiding-center potential energy
\begin{equation}
\Psi \;\equiv\; J\,\Omega \;+\; \epsilon\,\Psi_{1} \;+\; \epsilon^{2}\,\Psi_{2} \;+\; \cdots
\label{eq:Psi_def}
\end{equation}
is defined in terms of the gyroangle-independent scalar fields $\Psi_{n}$ ($n \geq 1$), which contain corrections due to magnetic-field nonuniformity. 

The guiding-center symplectic structure in Eq.~\eqref{eq:Gamma_H_gc}, on the other hand, is defined in terms of the Poincar\'{e}-Cartan one-form
\begin{eqnarray}
\Gamma_{\rm gc} & \equiv & \left( \frac{e}{\epsilon c}\,{\bf A} \;+\; \vb{\Pi} \right)\bdot\exd{\bf X} \;+\; \epsilon\,J\left(\exd\theta \;-\; {\bf R}\bdot\exd{\bf X}\right), 
\label{eq:Gamma_gc}
\end{eqnarray}
where the symplectic guiding-center momentum
\begin{equation}
\vb{\Pi} \;\equiv\; \sum_{n = 0}^{\infty}\epsilon^{n}\,\vb{\Pi}_{n} \;=\; p_{\|}\,\bhat \;+\; \epsilon\,\vb{\Pi}_{1} \;+\; \epsilon^{2}\,\vb{\Pi}_{2} + \cdots 
\label{eq:Pi_def}
\end{equation}
is expressed in terms of the gyroangle-independent vector fields $\vb{\Pi}_{n}$ ($n \geq 1$), which contain corrections due to magnetic-field nonuniformity. The presence of the gyrogauge vector ${\bf R}({\bf X})$ guarantees that the guiding-center one-form \eqref{eq:Gamma_gc} is gyrogauge-invariant \cite{RGL_1983}. 

Using Eqs.~\eqref{eq:Hamiltonian_gc} and \eqref{eq:Gamma_gc}, the guiding-center phase-space Lagrangian \eqref{eq:Gamma_H_gc} is thus expressed as
\begin{eqnarray}
\Lambda_{\rm gc} & = & \left[ \left(\frac{e}{\epsilon c}\,{\bf A} \;+\; p_{\|}\,\bhat \;-\; \epsilon\,J\,{\bf R}\right)\bdot\exd{\bf X} \;+\; \epsilon\,J\;\exd\theta \right] 
\label{eq:Lambda_gc} \\
 & - &\left( \frac{p_{\|}^{2}}{2m} + J\,\Omega \right) dt + \sum_{n=1}^{\infty}\epsilon^{n} \left( \Psi_{n}\frac{}{}dt - \vb{\Pi}_{n}\bdot\exd{\bf X} \right),
 \nonumber
 \end{eqnarray} 
 where higher-order corrections $(n \geq 1)$ are either contained in the guiding-center Hamiltonian $(\Psi_{n} \neq 0)$ or the guiding-center symplectic structure 
 $(\vb{\Pi}_{n} \neq 0)$. 
 
Guiding-center theories are said to be {\it equivalent} \cite{Brizard_Tronko_2012,Brizard_Tronko_2015} if they have the same definition of the guiding-center gyroaction $J$ but different definitions of the scalar field $\Psi$ and the vector field $\vb{\Pi}$. This equivalence class will be expressed at each order in terms of a relation involving the combination $\Psi_{n} - \Pi_{n\|}\;p_{\|}/m$, where $\Pi_{n\|} \equiv \bhat\bdot\vb{\Pi}_{n}$ denotes the parallel component of $\vb{\Pi}_{n}$. 

In a purely {\it Hamiltonian} representation ($\vb{\Pi}_{n} \equiv 0)$, the vector field $\vb{\Pi} \equiv p_{\|}\,\bhat$ is independent of the gyroaction $J$, while the scalar field $\Psi \equiv J\,\Omega + \epsilon\,\Psi_{1} + \epsilon^{2}\,\Psi_{2} + \cdots$ contains all the correction terms associated with the nonuniformity of the magnetic field. In a purely {\it symplectic} representation ($\Psi_{n} \equiv 0)$, on the other hand, the scalar field $\Psi \equiv J\,\Omega$ is independent of the parallel momentum $p_{\|}$, while the vector field $\vb{\Pi} = p_{\|}\,\bhat + \epsilon\,\vb{\Pi}_{1} + \cdots$ contains all the correction terms associated with the nonuniformity of the magnetic field. Our analysis will show that, while a purely Hamiltonian representation is possible at all orders, a purely symplectic representation is possible only at first order. We note that previous guiding-center Hamiltonian theories were constructed in a mixed representation (i.e., symplectic at first order and Hamiltonian at second order).

\subsubsection{Guiding-center Poisson bracket}

The guiding-center Poisson bracket obtained from the guiding-center Euler-Poincar\'{e} one-form \eqref{eq:Gamma_gc} by following the following inversion procedure. First, we construct the guiding-center Lagrange two-form $\vb{\omega}_{\rm gc} \equiv \exd\Gamma_{\rm gc}$. We note that the Lagrange component-matrix is invertible if the 
guiding-center Jacobian does not vanish \cite{footnote}
\begin{equation}
{\cal J}_{\rm gc} \equiv \sqrt{{\rm det}(\vb{\omega}_{\rm gc})} = \epsilon\;\bhat^{*}\bdot\left(\frac{e}{\epsilon\,c}\;{\bf B}^{*}\right) \equiv \frac{e}{c}\;B_{\|}^{**} \neq 0,
\label{eq:Jac_gc}
\end{equation}
where we use the following definitions
\begin{eqnarray}
{\bf B}^{*} & \equiv & \nabla\btimes\left[ {\bf A} \;+\; \frac{c}{e}\,\left( \epsilon\,\vb{\Pi} \;-\frac{}{}
\epsilon^{2}\;J\;{\bf R} \right) \right], \label{eq:Bstar_def} \\
\bhat^{*} & \equiv & \pd{\vb{\Pi}}{p_{\|}} \;=\; \bhat \;+\; \epsilon\;\pd{\vb{\Pi}_{1}}{p_{\|}} \;+\; \cdots, \label{eq:bstar_def} \\
{\bf R}^{*} & \equiv & {\bf R} \;-\; \epsilon^{-1}\;\pd{\vb{\Pi}}{J} \;=\; {\bf R} \;-\; \pd{\vb{\Pi}_{1}}{J} \;+\; \cdots, \label{eq:Rstar_def} \\
B_{\|}^{**} & \equiv & \bhat^{*}\bdot{\bf B}^{*} = \left( \bhat + \epsilon\pd{\vb{\Pi}_{1}}{p_{\|}} + \cdots \right)\bdot{\bf B}^{*}.
\label{eq:B||star_def}
\end{eqnarray}
Here, the fields ${\bf B}^{*}$ and $\bhat^{*}$ satisfy the identities $\nabla\bdot{\bf B}^{*} \equiv 0$, $\partial{\bf B}^{*}/\partial p_{\|} \equiv \epsilon\,(c/e)\,\nabla\btimes\bhat^{*}$, and $\partial{\bf B}^{*}/\partial J \equiv -\,\epsilon^{2}(c/e)\,\nabla\btimes{\bf R}^{*}$, which play an important role in the properties of the guiding-center Poisson bracket.

Next, we invert the guiding-center Lagrange matrix $\vb{\omega}_{\rm gc}$ to construct the guiding-center Poisson matrix with components 
$J_{\rm gc}^{\alpha\beta}$, such that $J_{\rm gc}^{\alpha\nu}\,(\omega_{\rm gc})_{\nu\beta} \equiv \delta^{\alpha}_{\;\beta}$. Lastly, we construct the guiding-center Poisson bracket $\{F,\; G\}_{\rm gc} \equiv (\partial F/\partial Z^{\alpha})\,J_{\rm gc}^{\alpha\beta}\,(\partial G/\partial Z^{\beta})$:
\begin{eqnarray}
\left\{ F,\frac{}{} G\right\}_{\rm gc} & = & \epsilon^{-1} \left( \pd{F}{\theta}\,\pd{G}{J} \;-\; \pd{F}{J}\,\pd{G}{\theta} \right) \nonumber \\
 &  &+\; \frac{{\bf B}^{*}}{B_{\|}^{**}}\bdot\left(\nabla^{*}F\;\pd{G}{p_{\|}} \;-\; \pd{F}{p_{\|}}\;\nabla^{*}G \right) \nonumber \\
 &  &-\; \frac{\epsilon\,c\bhat^{*}}{e\,B_{\|}^{**}}\bdot\nabla^{*}F\btimes\nabla^{*}G,
\label{eq:PB_gc_star}
\end{eqnarray}
where the modified gradient operator $\nabla^{*} \equiv \nabla + {\bf R}^{*}\partial/\partial\theta$ ensures gyrogauge-invariance \cite{Cary_Brizard_2009}. The derivation procedure of the guiding-center Poisson bracket \eqref{eq:PB_gc_star} guarantees that it satisfies the standard Poisson-bracket properties, while the guiding-center Jacobian \eqref{eq:Jac_gc} can be used to write Eq.~\eqref{eq:PB_gc_star} in phase-space divergence form
\begin{equation}
\left\{ F,\frac{}{} G\right\}_{\rm gc} \;=\; \frac{1}{{\cal J}_{\rm gc}}\;\pd{}{Z^{\alpha}}\left({\cal J}_{\rm gc}\;F\frac{}{} \left\{ Z^{\alpha},\;
G\right\}_{\rm gc}\right).
\label{eq:PBgc_div}
\end{equation} 

\subsection{Guiding-center Hamilton equations of motion}

The Hamiltonian guiding-center equations of motion 
\begin{eqnarray}
\frac{d_{\rm gc}Z^{\alpha}}{dt} & = & J_{\rm gc}^{\alpha\nu}\,(\omega_{\rm gc})_{\nu\beta}\,\frac{d_{\rm gc}Z^{\beta}}{dt} \nonumber \\
 & = & J_{\rm gc}^{\alpha\nu}\,\pd{H_{\rm gc}}{Z^{\nu}} \;\equiv\; \left\{ Z^{\alpha},\frac{}{} H_{\rm gc}\right\}_{\rm gc}
\end{eqnarray}
are expressed in terms of the guiding-center Hamiltonian \eqref{eq:Hamiltonian_gc} and the guiding-center Poisson bracket \eqref{eq:PB_gc_star} as
\begin{eqnarray}
\frac{d_{\rm gc}{\bf X}}{dt} & = & \left(\frac{p_{\|}}{m} + \pd{\Psi}{p_{\|}}\right)\;\frac{{\bf B}^{*}}{B_{\|}^{**}} 
\;+\; \frac{\epsilon\,c\bhat^{*}}{e\,B_{\|}^{**}}\btimes\nabla\Psi, 
\label{eq:Xdot_gc} \\
\frac{d_{\rm gc}p_{\|}}{dt} & = & -\;\frac{{\bf B}^{*}}{B_{\|}^{**}}\bdot\nabla\Psi,
\label{eq:pdot_gc} \\
\frac{d_{\rm gc}\theta}{dt} & = &\epsilon^{-1}\;\pd{\Psi}{J} \;+\; \frac{d_{\rm gc}{\bf X}}{dt}\bdot{\bf R}^{*},
\label{eq:thetadot_gc}
\end{eqnarray}
and 
\begin{equation}
\frac{d_{\rm gc}J}{dt} \;=\; -\,\epsilon^{-1}\,\pd{\Psi}{\theta} \;\equiv\; 0,
\label{eq:Jdot_gc}
\end{equation}
where the last equation follows from the effective guiding-center potential energy $\Psi$ being gyroangle-independent to all orders in $\epsilon$. We note that the Hamiltonian guiding-center equations of motion \eqref{eq:Xdot_gc}-\eqref{eq:pdot_gc} satisfy the guiding-center Liouville theorem
\begin{equation}
\nabla\bdot\left( B_{\|}^{**}\;\frac{d_{\rm gc}{\bf X}}{dt}\right) \;+\; \pd{}{p_{\|}}\left(B_{\|}^{**}\;\frac{d_{\rm gc}p_{\|}}{dt}\right) \;=\; 0,
\label{eq:gc_Liouville}
\end{equation}
which shows that the gyromotion action-angle dynamics, represented by Eqs.~\eqref{eq:thetadot_gc}-\eqref{eq:Jdot_gc}, is completely decoupled from the reduced guiding-center dynamics represented by Eqs.~\eqref{eq:Xdot_gc}-\eqref{eq:pdot_gc}.

In the guiding-center Hamilton equations \eqref{eq:Xdot_gc}-\eqref{eq:Jdot_gc}, the scalar field $\Psi$ appears explicitly, while the symplectic momentum
vector field $\vb{\Pi}$ appears implicitly in the guiding-center Poisson bracket through the vector fields ${\bf B}^{*}$, $\bhat^{*}$, and ${\bf R}^{*}$. The advantage of the Hamiltonian representation is that the guiding-center Poisson bracket is simplified by the choice $\vb{\Pi} = p_{\|}\,\bhat$, while the advantage of the symplectic representation is that the guiding-center Hamiltonian is simplified by the choice $\Psi = J\,\Omega$.

\section{\label{sec:Cgc}Consistent Guiding-center Transformation} 

The derivation of the guiding-center phase-space Lagrangian \eqref{eq:Gamma_H_gc} by Lie-transform phase-space Lagrangian perturbation method is based on a phase-space transformation from the local phase-space coordinates \eqref{eq:z0_def} to guiding-center coordinates $Z^{\alpha} = ({\bf X}, p_{\|}; J, \theta)$ generated by the vector fields 
$({\sf G}_{1}, {\sf G}_{2}, \cdots)$:
\begin{equation}
Z^{\alpha} \;=\; z_{0}^{\alpha} + \epsilon\,G_{1}^{\alpha} + \epsilon^{2}\,\left( G_{2}^{\alpha} + \frac{1}{2}\,{\sf G}_{1}\cdot\exd 
G_{1}^{\alpha}\right) + \cdots,
\label{eq:z_bar_z} 
\end{equation}
with its inverse defined as
\begin{equation}
z_{0}^{\alpha} \;=\; Z^{\alpha} - \epsilon\,G_{1}^{\alpha} - \epsilon^{2}\,\left( G_{2}^{\alpha} - \frac{1}{2}\,{\sf G}_{1}\cdot\exd 
G_{1}^{\alpha}\right) + \cdots.
\label{eq:zz_bar} 
\end{equation} 
In Eqs.~\eqref{eq:z_bar_z}-\eqref{eq:zz_bar}, the lowest-order guiding-center phase-space coordinates $z_{0}^{\alpha}$ are the local phase-space coordinates \eqref{eq:z0_def}, where ${\bf x}$ denotes the particle position, $p_{\|0} \equiv {\bf p}_0\bdot\bhat({\bf x})$ denotes the local parallel momentum as calculated from the magnetic unit vector $\bhat({\bf x})$ evaluated at the particle position ${\bf x}$, $J_{0} \equiv |{\bf p}_{\bot 0}|^{2}/2m\Omega({\bf x})$ denotes the lowest-order gyroaction, where ${\bf p}_{\bot 0} \equiv \bhat\btimes({\bf p}_0\btimes\bhat)$, and $\theta_{0}$ denotes the lowest-order gyroangle such that $\partial{\bf p}_{\bot 0}/\partial\theta_{0} = 
{\bf p}_{\bot 0}\btimes\bhat$. The Jacobian for the transformation to local phase-space coordinates $({\bf x},{\bf p}_0) \rightarrow ({\bf x}, p_{\|0}, J_{0}, \theta_{0})$ is ${\cal J}_{0} = m\Omega = e\,B/c$.

While the derivation of the guiding-center phase-space coordinates some freedom (e.g., choosing a Hamiltonian or a symplectic representation), we must ensure that these coordinates are chosen consistently. For this purpose, a set of constraints is introduced to verify consistency at each order. 

\subsection{Guiding-center Jacobian constraints}

The guiding-center Jacobian \eqref{eq:Jac_gc} associated with the phase-space transformation \eqref{eq:z_bar_z} is defined as
\begin{eqnarray}
{\cal J}_{\rm gc} & = & {\cal J}_{0} \;-\; \left. \pd{}{Z^{\alpha}}\right[ {\cal J}_{0}\frac{}{} \left(\epsilon\,G_{1}^{\alpha} \;+\frac{}{} 
\epsilon^{2}\,G_{2}^{\alpha} + \cdots\right) \nonumber \\
 &  &\left.-\; \frac{\epsilon^{2}}{2}\;G_{1}^{\alpha}\;\pd{}{Z^{\beta}}\left({\cal J}_{0}\frac{}{} G_{1}^{\beta} + \cdots\right) 
\;+\; \cdots \right] \nonumber \\
 & \equiv & {\cal J}_{0} \;+\; \epsilon\,{\cal J}_{1} \;+\; \epsilon^{2}\;{\cal J}_{2} \;+\; \cdots.
\label{eq:Jacobian_Lie}
\end{eqnarray}
Hence, at first and second orders, the components of the first and second order generating vector fields 
${\sf G}_{1}$ and ${\sf G}_{2}$ must satisfy the Jacobian constraints:
\begin{eqnarray}
\frac{{\cal J}_{1}}{{\cal J}_{0}} & = & \pd{\Pi_{1\|}}{p_{\|}} \;+\; \varrho_{\|}\,\tau \;\equiv\; -\;\frac{1}{{\cal J}_{0}} \pd{}{Z^{\alpha}}\left( 
{\cal J}_{0}\frac{}{} G_{1}^{\alpha}\right), \label{eq:Jac_1} \\
\frac{{\cal J}_{2}}{{\cal J}_{0}} & = & \pd{\Pi_{2\|}}{p_{\|}} + \varrho_{\|}\;\pd{\vb{\Pi}_{1}}{p_{\|}}\bdot\nabla\btimes\bhat + \frac{c\bhat}{eB}\bdot\nabla\btimes(\vb{\Pi}_{1} - J\,{\bf R}) \nonumber \\
 & \equiv & -\;\frac{1}{{\cal J}_{0}} \pd{}{Z^{\alpha}}\left( {\cal J}_{0}\; G_{2}^{\alpha} \;+\;\frac{1}{2}\;{\cal J}_{1}\,G_{1}^{\alpha} \right), 
\label{eq:Jac_2}
\end{eqnarray}
where $\varrho_{\|} \equiv p_{\|}/(m\Omega)$ and $\tau \equiv \bhat\bdot\nabla\btimes\bhat$.

\subsection{Guiding-center Hamiltonian constraints}

Another requirement for the guiding-center transformation \eqref{eq:z_bar_z} is that the definition of the guiding-center gyroaction $J$ must be unique, which leads to the following guiding-center Hamiltonian constraints \cite{Brizard_Tronko_2015}.

\subsubsection{First-order Hamiltonian constraint}

The second-order $(\epsilon^{2})$ Lie-transform perturbation analysis \cite{Brizard_Tronko_2015} yields the first-order $(\epsilon_{\rm B})$ guiding-center Hamiltonian constraint 
\begin{eqnarray}
\Psi_{1} \;-\; \frac{p_{\|}}{m}\;\Pi_{1\|} & \equiv & -\;\Omega\;\langle G_{1}^{J}\rangle \;-\; \frac{1}{2}\;J\,\Omega\;\varrho_{\|}\,\tau \nonumber \\
 & = & \frac{1}{2}\;J\,\Omega\;\varrho_{\|}\,\tau,
\label{eq:Ham_constraint_1}
\end{eqnarray}
where $\langle G_{1}^{J}\rangle \equiv -\,J\;\varrho_{\|}\,\tau$ is calculated at order $\epsilon^{3}$ in the Lie-transform perturbation analysis \cite{Brizard_Tronko_2015}. This first-order Hamiltonian constraint, of course, has an infinite number of solutions for $(\Pi_{1\|},\Psi_{1})$. One possible choice for $(\Pi_{1\|},\Psi_{1})$, for example, is $\Pi_{1\|} = \frac{1}{2}\,J\,\tau$ and $\Psi_{1} = J\,\Omega\;(\varrho_{\|}\tau)$, which allows the Ba\~{n}os parallel drift velocity $\partial\Psi_{1}/\partial p_{\|} = J\,\tau/m$ to be included in Eq.~\eqref{eq:Xdot_gc}. 

Here, we note that, since the right side of Eq.~\eqref{eq:Ham_constraint_1} is linear in $p_{\|}$, we may choose $\Psi_{1} \equiv 0$ without making $\Pi_{1\|}$ singular. We, therefore, choose the first-order symplectic representation
\begin{equation}
\left. \begin{array}{rcl}
\Psi_{1} & \equiv & 0 \\
 &  & \\
\Pi_{1\|} & \equiv & -\;\frac{1}{2}\,J\,\tau
\end{array} \right\},
\label{eq:Pi1||_def}
\end{equation} 
in accordance with standard guiding-center and gyrocenter Hamiltonian theories \cite{Cary_Brizard_2009,Brizard_Hahm_2007}.

\subsubsection{Second-order Hamiltonian constraint}

The third-order $(\epsilon^{3})$ Lie-transform perturbation analysis \cite{Brizard_Tronko_2015} yields the second-order $(\epsilon_{\rm B}^{2})$ guiding-center Hamiltonian constraint 
\begin{eqnarray}
\Psi_{2} \;-\; \frac{p_{\|}}{m}\;\Pi_{2\|} & \equiv & -\,\Omega\;\langle G_{2}^{J}\rangle \;+\; J\Omega\;\varrho_{\|}^{2} \left( \frac{1}{2}\,\tau^{2} -
\langle\alpha_{1}^{2}\rangle \right) \nonumber \\
 &  &+\; \vb{\Pi}_{1}\bdot{\bf v}_{\rm gc} \;-\; \frac{m}{2}\,|{\bf v}_{\rm gc}|^{2},
\label{eq:Ham_constraint_2_primitive}
\end{eqnarray}
where $\langle G_{2}^{J}\rangle$ is calculated at order $\epsilon^{4}$ \cite{Brizard_Tronko_2015}, we have defined the gyroangle-dependent scalar function
\begin{equation}
\alpha_{1} \;\equiv\; -\,\frac{1}{2} \left(\wh{\bot}\wh{\rho} + \wh{\rho}\wh{\bot}\right):\nabla\bhat 
\label{eq:alpha1_def}
\end{equation}
(where we use the rotating unit-vector basis $\wh{\bot} \equiv \wh{\rho}\btimes\bhat = \partial\wh{\rho}/
\partial\theta$), and ${\bf v}_{\rm gc}$ denotes the lowest-order guiding-center (perpendicular) drift velocity
\begin{equation}
{\bf v}_{\rm gc} \;\equiv\; \frac{\bhat}{m\Omega}\btimes\left( J\;\nabla\Omega \;+\; \frac{p_{\|}^{2}}{m}\;\vb{\kappa}\right),
\label{eq:vgc_def}
\end{equation}
where $\vb{\kappa} \equiv \bhat\bdot\nabla\bhat$ denotes the magnetic curvature. We now see that the perpendicular component $\vb{\Pi}_{1\bot}$ makes its appearance in Eq.~\eqref{eq:Ham_constraint_2_primitive}. 

When $\langle G_{2}^{J}\rangle$ is calculated at order $\epsilon^{4}$ in the Lie-transform perturbation analysis \cite{Brizard_Tronko_2015}, we find 
\begin{widetext}
\begin{eqnarray}
\langle G_{2}^{J}\rangle & = & \frac{J^{2}}{2m\Omega} \left[ \frac{\tau^{2}}{2} + \bhat\bdot\nabla\btimes{\bf R} - \langle\alpha_{1}^{2}
\rangle - \frac{\bhat}{2}\bdot\nabla\btimes(\bhat\btimes\nabla\ln B) \right] \;-\; \frac{J}{2}\,\varrho_{\|}^{2} \left[ \vb{\kappa}\bdot(3\,
\vb{\kappa} - \nabla\ln B) \;+\frac{}{} \nabla\bdot\vb{\kappa} - \tau^{2} \right].
\label{eq:G2_J_Ham}
\end{eqnarray} 
\end{widetext}
which, when inserted into Eq.~\eqref{eq:Ham_constraint_2_primitive}, yields the second-order  $(\epsilon_{\rm B}^{2})$ guiding-center Hamiltonian constraint \cite{Brizard_Tronko_2015}
\begin{eqnarray}
\Psi_{2} \;-\; \frac{p_{\|}}{m}\;\Pi_{2\|} & \equiv & J\,\Omega\left( \frac{J}{2\,m\Omega}\;\beta_{2\bot} \;+\; \frac{1}{2}\,\varrho_{\|}^{2}\;\beta_{2\|} \right) \nonumber \\
 &  &-\; \frac{p_{\|}^{2}}{2m}\;\left(\varrho_{\|}^{2}\frac{}{}|\vb{\kappa}|^{2}\right) \;+\; \vb{\Pi}_{1}\bdot{\bf v}_{\rm gc},
\label{eq:Hamiltonian_constraint_2}
\end{eqnarray}
where the gyroangle-independent scalar fields $\beta_{2\bot}({\bf X})$ and $\beta_{2\|}({\bf X})$ are defined as
\begin{eqnarray}
\beta_{2\bot} & = & -\,\frac{1}{2}\,\tau^{2} \;-\; \bhat\bdot\nabla\btimes{\bf R} \;+\; \langle\alpha_{1}^{2}\rangle \;-\; \left|\bhat\btimes\nabla\ln B\right|^{2} \nonumber \\
 &  &+\; \frac{1}{2}\;\bhat\bdot\nabla\btimes\left(\bhat\btimes\nabla\ln B\right),
\label{eq:beta2_perp} \\
\beta_{2\|} & = & -\,2\;\langle\alpha_{1}^{2}\rangle \;-\; 3\;\vb{\kappa}\bdot\left(\nabla\ln B \;-\frac{}{} \vb{\kappa}\right) \;+\; 
\nabla\bdot\vb{\kappa},
\label{eq:beta2_par}
\end{eqnarray}
with the definitions \cite{Brizard_Tronko_2015}
\begin{equation}
\bhat\bdot\nabla\btimes{\bf R} \;=\; \frac{1}{2}\;\nabla\bdot\left[ \vb{\kappa} \;-\frac{}{} \bhat\;(\nabla\bdot\bhat)\right],
\end{equation}
and
\begin{equation}
\langle \alpha_{1}^{2}\rangle \;=\; \frac{1}{2}\,\bhat\bdot\nabla\btimes{\bf R} \;+\; \frac{1}{8} \left[ \tau^{2} \;+\; \left(\nabla\bdot\bhat\right)^{2}
\right].
\label{eq:alpha1_square}
\end{equation}
The last term in Eq.~\eqref{eq:Hamiltonian_constraint_2}, which involves $\vb{\Pi}_{1\bot}$, is ignored in all previous works since it was previously assumed that 
$\vb{\Pi}_{1\bot} = 0$.

We now note that, in contrast to first-order guiding-center Hamiltonian constraint \eqref{eq:Ham_constraint_1}, the right side of Eq.~\eqref{eq:Hamiltonian_constraint_2} contains terms that are constant, quadratic, and quartic in $p_{\|}$. Hence, since Eq.~\eqref{eq:beta2_perp} shows that $\beta_{2\bot} \neq 0$, we cannot choose $\Psi_{2} = 0$ without making $\Pi_{2\|}$ singular in $p_{\|}$, i.e., a purely symplectic representation is no longer possible at second order. 

\subsection{Previous second-order Hamiltonian representations}

In order to compare our results with the results presented in Refs.~\cite{Parra_Calvo_2011,Burby_SQ_2013}, going back to Littlejohn's work 
\cite{RGL_1983}, we choose $\Pi_{2\|} \equiv 0$ and temporarily set $\vb{\Pi}_{1\bot} \equiv 0$ in Eq.~\eqref{eq:Hamiltonian_constraint_2}. Hence, with these simplifying assumptions, our work agrees with the second-order guiding-center Hamiltonian of Burby, Squire, and Qin (BSQ) \cite{Burby_SQ_2013}:
\begin{equation}
\Psi_{2(TB)} \;=\; \Psi_{2(BSQ)} \;=\; \Psi_{2(PC)} \;+\; \frac{d_{0}\langle\sigma_{3(PC)}\rangle}{dt},
\label{eq:Psi2_BT_BSQ_PC}
\end{equation}
while it agrees with the second-order guiding-center Hamiltonian of Parra and Calvo (PC) \cite{Parra_Calvo_2011} only up to the lowest-order guiding-center time derivative of the gyroangle-independent third-order gauge function 
\begin{equation} 
\langle\sigma_{3(PC)}\rangle \;=\; \frac{1}{2}\,J\;\varrho_{\|}\,(\nabla\bdot\bhat) \;\equiv\; \frac{d_{0}}{dt}\left(\frac{J}{2\Omega}\right)
\label{eq:PC_gauge}
\end{equation}
in the same manner discussed in Ref.~\cite{PCBSQ_2014}, where the lowest-order guiding-center time derivative is defined as $d_{0}/dt \equiv (p_{\|}/m)\bhat\bdot\nabla + 
J\Omega\,(\nabla\bdot\bhat)\,\partial/\partial p_{\|}$. 

We note that the guiding-center phase-space Lagrangian $\Lambda_{\rm gc(PC)} \equiv L_{\rm gc(PC)}\,dt$ of Parra and Calvo \cite{Parra_Calvo_2011}  differs from the other two guiding-center phase-space Lagrangians $L_{\rm gc(BSQ)} = L_{\rm gc(TB)}$ by an exact time derivative $d_{0}\langle\sigma_{3(PC)}\rangle/dt$. Since two Lagrangians $L({\bf q},\dot{\bf q},t)$ and $L^{\prime}({\bf q},\dot{\bf q},t)$ on configuration space ${\bf q}$ that differ by an exact time derivative $L^{\prime} \equiv L + dF/dt$ yield the same Euler-Lagrange equations \cite{Brizard_Lag} for any function $F({\bf q},t)$, the Lagrangians of Parra and Calvo \cite{Parra_Calvo_2011} and Burby, Squire, and Qin \cite{Burby_SQ_2013} are said to be equivalent \cite{footnote_2}. 

Lastly, in our previous work \cite{Brizard_Tronko_2012}, where $\vb{\Pi}_{1\bot} \equiv 0$ was assumed, we selected the following mixed representation: the second-order symplectic term $\Pi_{2\|}(p_{\|},J,{\bf X}) = \frac{1}{2}\,p_{\|}\,[ \varrho_{\|}^{2}|\vb{\kappa}|^{2} - (J/m\,\Omega)\,\beta_{2\|}]$, and the second-order Hamiltonian term 
$\Psi_{2}(J,{\bf X}) \equiv (J^{2}/2m)\,\beta_{2\bot}$, which follows from Eq.~\eqref{eq:Hamiltonian_constraint_2}, was not included in Ref.~\cite{Brizard_Tronko_2012}.

\subsection{Guiding-center transformation}

The full Lie-transform perturbation analysis leading to the present higher-order guiding-center Hamiltonian theory will be presented elsewhere \cite{Brizard_Tronko_2015}. Here, we summarize the guiding-center phase-space transformation $z_{0}^{\alpha} \equiv ({\bf x},p_{\|0},J_{0},\theta_{0}) \rightarrow Z^{\alpha} \equiv ({\bf X}, p_{\|},J,\theta)$,
defined in Eq.~\eqref{eq:z_bar_z}  by the first-order generating vector-field components
\begin{eqnarray}
G_{1}^{\bf x} & = & -\;\vb{\rho}_{0} \;\equiv\; {\bf p}_{\bot 0}\btimes\bhat/m\Omega, \label{eq:G1_x} \\
G_{1}^{p_{\|}} & = & -\;p_{\|0}\;\vb{\rho}_{0}\bdot\vb{\kappa} \;+\; J_{0} \left( \tau \;+\; \alpha_{1} \right), \label{eq:G1_p} \\
G_{1}^{J} & = & \vb{\rho}_{0}\bdot\left( J_{0}\;\nabla\ln B + \frac{p_{\|0}^{2}\,\vb{\kappa}}{m\Omega}\right) \nonumber \\
 &  &-\; J_{0}\,\varrho_{\|0}\,(\tau + \alpha_{1}), \label{eq:G1_J} \\
G_{1}^{\theta} & = & \pd{\vb{\rho}_{0}}{\theta_{0}}\vb{\cdot}\left(\nabla\ln B + \frac{p_{\|0}^{2}\vb{\kappa}}{2\,mJ_{0}\Omega}\right) \nonumber \\
 &  &-\; \vb{\rho}_{0}\bdot{\bf R} \;+\; \varrho_{\|0}\;\alpha_{2}, \label{eq:G1_theta} 
\end{eqnarray}
where $\alpha_{1} \equiv \partial\alpha_{2}/\partial\theta_{0}$, and the second-order generating vector-field components
\begin{eqnarray}
G_{2}^{\bf x} & = & \left( 2\,\varrho_{\|0}\;\pd{\vb{\rho}_{0}}{\theta_{0}}\bdot\vb{\kappa} \;+\; \frac{J_{0}\,\alpha_{2}}{m\Omega}\right) \bhat \;-\; \vb{\Pi}_{1}\btimes\frac{\bhat}{m\Omega} \nonumber \\
 &  &+\; \frac{1}{2} \left[ \frac{p_{\|0}^{2}}{m\Omega}\;(\vb{\rho}_{0}\bdot\vb{\kappa}) + J_{0}\,\varrho_{\|0}\;(3 \tau - \alpha_{1}) \right] \pd{\vb{\rho}_{0}}{J_{0}} \label{eq:G2_x} \\
 &  &+\; \frac{1}{2} \left[ \varrho_{\|0}\,\alpha_{2} + \pd{\vb{\rho}_{0}}{\theta_{0}}\bdot \left( \nabla\ln B + \frac{p_{\|0}^{2}\,\vb{\kappa}}{2m\Omega\,J_{0}}\right) \right] 
 \pd{\vb{\rho}_{0}}{\theta_{0}}, \nonumber \\
G_{2}^{p_{\|}} & = & p_{\|0}\;\vb{\kappa}\frac{}{}\bdot G_{2}^{\bf x} \;+\; \bhat\bdot\left[D_{1}^{2}({\bf P}_{3}) + \nabla\sigma_{3} - \vb{\Pi}_{2}\right],
\label{eq:G2_p} \\
G_{2}^{J} & = & -\; \frac{1}{\Omega} \left( \Psi_{2} \;-\; \frac{p_{\|}}{m}\;\Pi_{2\|} \right) - \varrho_{\|}\bhat\bdot
\left[D_{1}^{2}({\bf P}_{3}) + \nabla\sigma_{3}\right] \nonumber \\
 &  &- G_{2}^{\bf x}\bdot\left(J_{0}\nabla\ln B + \frac{p_{\|0}^{2}\,\vb{\kappa}}{m\,\Omega} \right), \label{eq:G2_J}
\end{eqnarray}
where ${\bf P}_{3} \equiv \frac{1}{2}\,p_{\|0}\bhat + \frac{1}{3}\,{\bf p}_{\bot 0}$. We note that the spatial component
\begin{eqnarray}
G_{3}^{\bf x} & = & G_{3\|}^{\bf x}\;\bhat \;+\; G_{2\|}^{\bf x}\;\left(\varrho_{\|0}\frac{}{}\nabla\btimes\bhat\right) \;-\; G_{2}^{\bf x}\;
\left(\varrho_{\|0}\frac{}{}\tau \right) \nonumber \\
 &  &-\; \frac{c\bhat}{eB}\btimes\left[D_{1}^{2}({\bf P}_{3}) + \nabla\sigma_{3} - \vb{\Pi}_{2}\right],
\label{eq:G3_x}
\end{eqnarray}
which is determined at third order \cite{Brizard_Tronko_2015}, is not needed in this Section and the remaining components $G_{3\|}^{\bf x}$ and $G_{2}^{\theta}$, which are determined at fourth order, are not needed in what follows. In the expressions above, we used the definition
\begin{eqnarray*}
D_{1}(\cdots) & \equiv & \left(G_{1}^{p_{\|}}\pd{}{p_{\|0}} + G_{1}^{J}\pd{}{J_{0}} + G_{1}^{\theta}\pd{}{\theta_{0}}\right)(\cdots) \\
 &  &+\; \vb{\rho}_{0}\btimes \nabla\btimes(\cdots),
\end{eqnarray*}
and the gyroangle-dependent gauge function
\begin{equation}
\sigma_{3} \;\equiv\; -\,\frac{1}{3}\,p_{\|0}\;G_{2\|}^{\bf x}
\label{eq:sigma3_def}
\end{equation}
appearing in the third-order Lie-transform perturbation analysis \cite{Brizard_Tronko_2015}.

\subsection{Push-forward Lagrangian Constraints} 

The second-order guiding-center Hamiltonian constraint \eqref{eq:Hamiltonian_constraint_2} leads to a complex expression whose interpretation for 
$\Psi_{2}$ and $\Pi_{2\|}$ may be difficult to obtain. For this purpose, we wish to explore a new perturbation approach to guiding-center Hamiltonian theory. 

We begin with the following remark for the phase-space Lagrangian formulation of single-particle dynamics in a potential $U({\bf x})$, where the particle position ${\bf x}$ and its velocity ${\bf v}$ are viewed as independent phase-space coordinates. From the phase-space Lagrangian 
\[ L({\bf x},{\bf v};\dot{\bf x},\dot{\bf v}) = \left(\frac{e}{c}{\bf A} + m{\bf v}\right)\bdot\dot{\bf x} - \left(\frac{m}{2}\,|{\bf v}|^{2} + e\Phi\right), \]
we first obtain the Euler-Lagrange equation for ${\bf x}$: $m\,d{\bf v}/dt = e\,{\bf E} + {\bf v}\btimes e\,{\bf B}/c$. Since the phase-space Lagrangian is independent of $d{\bf v}/dt$, however, the Euler-Lagrange equation for ${\bf v}$ yields the Lagrangian constraint
\begin{equation}
\pd{L}{\bf v} \;=\; m\;\left(\frac{d{\bf x}}{dt} \;-\; {\bf v}\right) \;\equiv\; 0.
\label{eq:Lag_constraint}
\end{equation}
Hence, the guiding-center transformation of the particle velocity ${\bf v}$ is constrained to be also expressed in terms of the guiding-center transformation of $d{\bf x}/dt$.

We would now like to obtain the guiding-center version of the Lagrangian constraint \eqref{eq:Lag_constraint}:
\begin{equation}
{\sf T}_{\rm gc}^{-1}{\bf p}_{0} \;=\; m\,{\sf T}_{\rm gc}^{-1}\left(\frac{d{\bf x}}{dt}\right) \;\equiv\; {\bf P}_{\rm gc}.
\label{eq:gc_Lag}
\end{equation} 
First, using the functional definition for $d_{\rm gc}/dt$: 
\begin{equation}
\frac{d_{\rm gc}}{dt} \;\equiv\; {\sf T}_{\rm gc}^{-1}\left(\frac{d}{dt}\;{\sf T}_{\rm gc}\right),
\label{eq:d_gc_def}
\end{equation}
we introduced in Eq.~\eqref{eq:gc_Lag} the guiding-center particle-momentum
\begin{equation}
{\bf P}_{\rm gc} \;=\; m\,\frac{d_{\rm gc}}{dt}\left({\sf T}_{\rm gc}^{-1}{\bf x} \right) = m\,\frac{d_{\rm gc}{\bf X}}{dt} + m\,
\frac{d_{\rm gc}\vb{\rho}_{\rm gc}}{dt},
\label{eq:gcLc_id} 
\end{equation}
which is expressed as the sum of the guiding-center velocity 
\[ \frac{d_{\rm gc}{\bf X}}{dt} \;=\; \frac{d_{0}{\bf X}}{dt} + \epsilon\,\frac{d_{1}{\bf X}}{dt} + \cdots \;=\; \frac{p_{\|}}{m}\;\bhat + 
\epsilon\,{\bf v}_{\rm gc} + \cdots \]
and the guiding-center displacement velocity 
\[ \frac{d_{\rm gc}\vb{\rho}_{\rm gc}}{dt} = \epsilon^{-1}\;\pd{\Psi}{J}\;\pd{\vb{\rho}_{\rm gc}}{\theta} + \frac{d_{\rm gc}{\bf X}}{dt}\bdot
\nabla^{*}\vb{\rho}_{\rm gc} + \frac{d_{\rm gc}p_{\|}}{dt}\,\pd{\vb{\rho}_{\rm gc}}{p_{\|}}, \]
where
\[ \frac{d_{\rm gc}p_{\|}}{dt} \;=\; \frac{d_{0}p_{\|}}{dt} + \epsilon\,\frac{d_{1}p_{\|}}{dt} + \cdots \;=\; J\,\Omega\;\left(\nabla\bdot\bhat\right) + \cdots. \]
Here, the guiding-center displacement is expanded as
\begin{equation}
\vb{\rho}_{\rm gc} \;\equiv\; {\sf T}_{\rm gc}^{-1}{\bf x} \;-\; {\bf X} \;=\; \epsilon\,\vb{\rho}_{0} \;+\; \epsilon^{2}\;
\vb{\rho}_{1} \;+\; \epsilon^{3}\,\vb{\rho}_{2} + \cdots,
\label{eq:rho_gc}
\end{equation}
where the higher-order gyroradius corrections are
\begin{eqnarray}
\vb{\rho}_{1} & = & -\;G_{2}^{\bf x} \;-\; \frac{1}{2}\;{\sf G}_{1}\cdot\exd\vb{\rho}_{0}, \label{eq:rho_1} \\
\vb{\rho}_{2} & = & -\;G_{3}^{\bf x} - {\sf G}_{2}\cdot\exd\vb{\rho}_{0} + \frac{1}{6}\,{\sf G}_{1}\cdot
\exd({\sf G}_{1}\cdot\exd\vb{\rho}_{0}). \label{eq:rho_2}
\end{eqnarray}
We note that, in general, we find $\langle\vb{\rho}_{n}\rangle \neq 0$ and $\vb{\rho}_{n}\bdot\bhat \neq 0$ for $n \geq 1$. 

\subsubsection{First-order Lagrangian constraint}

The first-order Lagrangian constraints on the components $(G_{1}^{p_{\|}}, G_{1}^{J}, G_{1}^{\theta})$ are expressed as
\begin{equation}
G_{1}^{p_{\|}}\;\bhat + G_{1}^{J}\,\pd{{\bf p}_{\bot 0}}{J} + G_{1}^{\theta}\,\pd{{\bf p}_{\bot 0}}{\theta} - \vb{\rho}_{0}\bdot\nabla{\bf p}_{0} + 
{\bf P}_{{\rm gc}1} \;\equiv\; 0,
\label{eq:gcLc_1}
\end{equation}
where
\[ {\bf P}_{{\rm gc}1} \equiv m\,\frac{d_{1}{\bf X}}{dt} \;+\; m\,\left(\frac{d_{\rm gc}\vb{\rho}_{\rm gc}}{dt}\right)_{1}, \]
with 
\[ \left(\frac{d_{\rm gc}\vb{\rho}_{\rm gc}}{dt}\right)_{1} \;\equiv\; \Omega\;\pd{\vb{\rho}_{1}}{\theta} \;+\; \frac{d_{0}\vb{\rho}_{0}}{dt}, \]
and
\[ \frac{d_{0}\vb{\rho}_{0}}{dt} \;\equiv\; \frac{p_{\|}}{m}\bhat\bdot\left[\nabla\vb{\rho}_{0} \;+\; \left({\bf R} + \pd{\vb{\Pi}_{1}}{J}\right)
\pd{\vb{\rho}_{0}}{\theta}\right]. \]
Using the identity
\[ \Psi_{1} \;=\; \left\langle {\bf p}_{\bot 0}\bdot\left(\frac{d_{\rm gc}\vb{\rho}_{\rm gc}}{dt}\right)_{1}\right\rangle \;\equiv\; 0, \]
which follows from the first-order symplectic representation \eqref{eq:Pi1||_def}, Eq.~\eqref{eq:gcLc_1} yields the same condition used in the first-order Hamiltonian constraint \eqref{eq:Ham_constraint_1}:
\begin{equation}
\langle G_{1}^{J}\rangle \;=\; \left\langle \vb{\rho}_{0}\bdot\nabla{\bf p}_{0}\bdot\pd{\vb{\rho}_{0}}{\theta}\right\rangle \;=\; -\;J\,\varrho_{\|}\tau,
\label{eq:G1_J_ave}
\end{equation}
which is calculated at order $\epsilon^{3}$ in the Lie-transform perturbation analysis \cite{Brizard_Tronko_2015}.

\subsubsection{Second-order Lagrangian constraint}

The second-order components $(G_{2}^{p_{\|}}, G_{2}^{J}, G_{2}^{\theta})$ are also constrained by the second-order Lagrangian constraint
\begin{eqnarray}
 &  &G_{2}^{p_{\|}}\;\bhat \;+\; G_{2}^{J}\,\pd{{\bf p}_{\bot 0}}{J} \;+\; G_{2}^{\theta}\,\pd{{\bf p}_{\bot 0}}{\theta} \;+\; G_{2}^{\bf x}\bdot\nabla{\bf p}_{0} \nonumber \\
 &  &\;-\; \frac{1}{2}\;{\sf G}_{1}\cdot\exd\left( {\sf G}_{1}\cdot\exd{\bf p}_{0} \right) \;+\; {\bf P}_{{\rm gc}2} \;\equiv\; 0,
\label{eq:gcLc_2}
\end{eqnarray}
where
\[ {\bf P}_{{\rm gc}2} \;\equiv\; m\,\frac{d_{2}{\bf X}}{dt} + m\,\left(\frac{d_{\rm gc}\vb{\rho}_{\rm gc}}{dt}\right)_{2} \]
with
\[ \left(\frac{d_{\rm gc}\vb{\rho}_{\rm gc}}{dt}\right)_{2} \equiv \Omega\;\pd{\vb{\rho}_{2}}{\theta} + \pd{\Psi_{2}}{J}\,\pd{\vb{\rho}_{0}}{\theta} + 
\frac{d_{1}{\bf X}}{dt}\bdot\nabla_{0}^{*}\vb{\rho}_{0} + \frac{d_{0}\vb{\rho}_{1}}{dt}, \]
and 
\begin{eqnarray*} 
\frac{d_{0}\vb{\rho}_{1}}{dt} & = & \frac{p_{\|}}{m}\bhat\bdot\left[\nabla\vb{\rho}_{1} \;+\; \left({\bf R} + \pd{\vb{\Pi}_{1}}{J}\right)
\pd{\vb{\rho}_{1}}{\theta}\right] \\
 &  &+\; \left[ J\,\Omega\;\left(\nabla\bdot\bhat\right)\right]\;\pd{\vb{\rho}_{1}}{p_{\|}}.
\end{eqnarray*}
In particular, the Lagrangian constraint on $\langle G_{2}^{J}\rangle$ yields
\begin{eqnarray}
\langle G_{2}^{J}\rangle & = & -\;\left\langle G_{2}^{\bf x}\bdot\nabla{\bf p}_{0}\bdot\pd{\vb{\rho}_{0}}{\theta}\right\rangle - m\,\left\langle\left(
\frac{d_{\rm gc}\vb{\rho}_{\rm gc}}{dt}\right)_{2}\bdot\pd{\vb{\rho}_{0}}{\theta}\right\rangle \nonumber \\
 &  &+\; \frac{1}{2}\;\left\langle\left[{\sf G}_{1}\cdot\exd\frac{}{}\left( {\sf G}_{1}\cdot\exd{\bf p}_{0} \right)\right]\bdot\pd{\vb{\rho}_{0}}{\theta}\right\rangle,
\label{eq:G2_J_ave}
\end{eqnarray}
which yields the same result as Eq.~\eqref{eq:G2_J_Ham} obtained at order $\epsilon^{4}$ in the Lie-transform perturbation analysis \cite{Brizard_Tronko_2015}.

\subsubsection{Lagrangian constraint on the guiding-center Hamiltonian}

The generating-field components \eqref{eq:G1_x}-\eqref{eq:G2_J} were shown to satisfy the guiding-center Lagrangian constraints 
\eqref{eq:gcLc_1}-\eqref{eq:gcLc_2}. This means that the guiding-center Hamiltonian
\begin{eqnarray}
H_{\rm gc} & \equiv & \frac{m}{2} \left\langle \left|\frac{d_{\rm gc}{\bf X}}{dt} \;+\; \frac{d_{\rm gc}\vb{\rho}_{\rm gc}}{dt}\right|^{2}\right\rangle, 
\label{eq:Ham_gc_final}
\end{eqnarray}
can also be expressed in terms of guiding-center velocity $d_{\rm gc}{\bf X}/dt$ and the guiding-center displacement velocity 
$d_{\rm gc}\vb{\rho}_{\rm gc}/dt$. In the second-order Hamiltonian representation $(\Pi_{2\|} \equiv 0)$, the Lagrangian constraint of the guiding-center Hamiltonian \eqref{eq:Ham_gc_final} implies that
\begin{eqnarray}
\Psi_{2} \equiv \epsilon^{-2} \left[ \frac{m}{2} \left\langle \left|\frac{d_{\rm gc}{\bf X}}{dt} + \frac{d_{\rm gc}\vb{\rho}_{\rm gc}}{dt}\right|^{2}\right\rangle \;-\; \left( \frac{p_{\|}^{2}}{2m} + J\,\Omega \right)  \right],
\end{eqnarray}
which is identical to Eq.~\eqref{eq:Hamiltonian_constraint_2} (with $\Pi_{2\|} \equiv 0)$.

\section{\label{sec:gcPol_TCM}Guiding-center Polarization and Toroidal Canonical Momentum} 

There is now well-established connection between polarization and the conservation of toroidal canonical momentum in an axisymmetric magnetic field.
We now show how $\vb{\Pi}_{1\bot}$, which was originally chosen by Littlejohn \cite{RGL_1983} to be zero, can be determined by requiring that the guiding-center transformation \eqref{eq:z_bar_z} yields the guiding-center polarization obtained by Pfirsch \cite{Pfirsch_1984} and Kaufman 
\cite{Kaufman_1986}. We will also show that the {\it polarization} term $\vb{\Pi}_{1\bot}$ leads to a more transparent guiding-center representation of the toroidal canonical angular momentum in axisymmetric magnetic geometry.

\subsection{Guiding-center polarization}

The guiding-center transformation \eqref{eq:z_bar_z} can be used to calculate polarization and magnetization effects associated with the guiding-center displacement $\vb{\rho}_{\rm gc}$, defined by Eq.~\eqref{eq:rho_gc}. 

Since the dipole contribution to the guiding-center polarization \cite{Brizard_2013} involves the gyroangle-averaged displacement $\langle\vb{\rho}_{\rm gc}\rangle = \epsilon^{2}\,\langle\vb{\rho}_{1}\rangle + \cdots$ (since $\langle\vb{\rho}_{0}\rangle \equiv 0$), we begin with the gyroangle-averaged first-order displacement calculated from Eq.~\eqref{eq:rho_1} \cite{Brizard_Tronko_2011}:
\begin{eqnarray}
\langle\vb{\rho}_{1}\rangle & = & -\;\frac{J}{m\Omega} \left[ \frac{1}{2}\,(\nabla\bdot\bhat)\,\bhat \;+\; \frac{3}{2}\,\nabla_{\bot}\ln B\right] \nonumber \\
 &  &-\; \varrho_{\|}^{2}\,\vb{\kappa} \;+\; \vb{\Pi}_{1}\btimes\frac{\bhat}{m\Omega} \nonumber \\
 & \equiv & -\,\frac{1}{m\Omega}\,\left( J\;\nabla_{\bot}\ln B + \frac{p_{\|}^{2}\,\vb{\kappa}}{m\,\Omega}\right) + \nabla\bdot\left(\left\langle
\frac{\vb{\rho}_{0}\vb{\rho}_{0}}{2}\right\rangle\right) \nonumber \\
 &  &+\; \left( \frac{J}{2} \;\bhat\btimes\vb{\kappa} \;+\; \vb{\Pi}_{1}\right)\btimes\frac{\bhat}{m\Omega},
\label{eq:rho1_ave}
\end{eqnarray}
where we used Eqs.~\eqref{eq:G1_x}-\eqref{eq:G2_x}, with 
\begin{eqnarray*}
\nabla\bdot\left(\left\langle\frac{\vb{\rho}_{0}\vb{\rho}_{0}}{2}\right\rangle\right) & = & \nabla\bdot\left[ \frac{J}{2\,m\Omega}\;\left({\bf I} - \bhat\bhat\right) \right] \\
 & = & -\,\frac{J}{2\,m\Omega} \left[\vb{\kappa} + \nabla_{\bot}\ln B \;+\; (\nabla\bdot\bhat)\,\bhat \right].
\end{eqnarray*}
Next, the guiding-center polarization density is defined as the first-order expression \cite{Brizard_2013} 
\begin{eqnarray}
\vb{\pi}_{\rm gc}^{(1)} & \equiv & e\,\langle\vb{\rho}_{1}\rangle \;-\; e\;\nabla\bdot\left(\left\langle\frac{\vb{\rho}_{0}\vb{\rho}_{0}}{2}\right\rangle\right) \nonumber \\
 & = & -\,\frac{e}{m\Omega}\,\left( J\;\nabla_{\bot}\ln B \;+\; \frac{p_{\|}^{2}\,\vb{\kappa}}{m\,\Omega}\right) \nonumber \\
 &  &+\; \left( \frac{J}{2} \;\bhat\btimes\vb{\kappa} \;+\; \vb{\Pi}_{1}\right)\btimes\frac{c\bhat}{B},
\label{eq:gc_pol_def}
\end{eqnarray}
which yields the Pfirsch-Kaufman formula \cite{Pfirsch_1984,Kaufman_1986}
\begin{equation}
\vb{\pi}_{\rm gc}^{(1)} \;\equiv\; e\;\bhat\btimes \frac{1}{\Omega}\frac{d_{1}{\bf X}}{dt} \;=\; e\;\bhat\btimes\frac{{\bf v}_{\rm gc}}{\Omega},
\label{gc_pol_PK}
\end{equation}
only if we use the definition \cite{Brizard_2013}
\begin{equation}
\vb{\Pi}_{1\bot} \;\equiv\; -\;\frac{J}{2}\;\bhat\btimes\vb{\kappa}.
\label{eq:Pi1_perp_choice}
\end{equation}
Hence, by combining with the condition \eqref{eq:Pi1||_def}, $\Pi_{1\|} \equiv \bhat\bdot\vb{\Pi}_{1} = -\,\frac{1}{2}\,J\,\tau$, we find
\begin{equation}
\vb{\Pi}_{1} \;=\; -\;\frac{J}{2} \left( \tau\;\bhat \;+\frac{}{} \bhat\btimes\vb{\kappa} \right) \;=\; -\;\frac{J}{2}\;\nabla\btimes\bhat,
\label{eq:Pi1_final}
\end{equation}
and, hence, the guiding-center vector \eqref{eq:bstar_def} becomes $\bhat^{*} = \bhat + {\cal O}(\epsilon^{2})$, since $\partial\vb{\Pi}_{1}/\partial p_{\|} = 0$. 

We note that the Pfirsch-Kaufman formula \eqref{gc_pol_PK} yields a guiding-center moving-electric-dipole correction $\vb{\mu}_{\rm gc}^{(E)} \equiv p_{\|}{\bf v}_{\rm gc}/B$ to the intrinsic guiding-center magnetic-dipole moment $\vb{\mu}_{\rm gc}^{(B)} \equiv -\,\mu\,\bhat$. Since the total guiding-center magnetic-dipole moment can be obtained from the Lagrangian variational expression \cite{CRPW_1986,Tronci}
\begin{eqnarray}
\vb{\mu}_{\rm gc} & = & -\;\mu\;\bhat \;+\; \epsilon\;p_{\|}\;{\bf v}_{\rm gc}/B \label{eq:mu_gc_total} \\
 & \equiv & \fd{}{\bf B}\left[ \left(\frac{e}{\epsilon c}\,{\bf A} + p_{\|}\,\bhat\right)\bdot\dot{\bf X} \;-\; \left( \mu\,B \;+\; \frac{p_{\|}^{2}}{2m}\right) \right],
 \nonumber
 \end{eqnarray}
 where we used $\delta\bhat/\delta{\bf B} = ({\bf I} - \bhat\bhat)/B$ and $\delta B/\delta{\bf B} = \bhat$, we again conclude that the term $\vb{\Pi}_{1\bot}$, defined by 
 Eq.~\eqref{eq:Pi1_perp_choice}, cannot be ignored in guiding-center theory if polarization effects are to be accounted for correctly.

Lastly, the guiding-center phase-space Lagrangian is expressed as
\begin{eqnarray}
\Gamma_{\rm gc} & = & \left( \frac{e}{\epsilon\,c}\;{\bf A} \;+\; p_{\|}\,\bhat \;-\; \frac{\epsilon}{2}\;J\,\nabla\btimes\bhat \right)\bdot\exd{\bf X} 
\nonumber \\
 &  &+\; \epsilon\,J\;\left(\exd\theta \;-\frac{}{} {\bf R}\bdot\exd{\bf X}\right),
\label{eq:Gamma_gc_primitive}
\end{eqnarray}
when terms up to first order in magnetic-field nonuniformity are retained. In Eq.~\eqref{eq:Gamma_gc_primitive}, we have retained the guiding-center polarization contribution 
\eqref{eq:Pi1_perp_choice} to $\vb{\Pi}_{1} \equiv -\,\frac{1}{2}\,J\;\nabla\btimes\bhat$. We now show that this polarization correction yields a more transparent expression for the guiding-center toroidal canonical momentum up to second order in $\epsilon$ (i.e., first order in magnetic-field nonuniformity).

\subsection{Guiding-center toroidal canonical angular momentum}

We now construct the guiding-center representation for the toroidal canonical angular momentum in axisymmetric magnetic geometry, for which it is an exact constant of motion. Here, we represent an axisymmetric magnetic field
\begin{equation}
{\bf B} \;=\; B_{\varphi}(\psi)\;\nabla\varphi \;+\; \nabla\varphi\btimes\nabla\psi,
\label{eq:B_axis}
\end{equation}
where $\varphi$ denotes the toroidal angle and $\psi$ denotes the magnetic flux on which magnetic-field lines lie (i.e., ${\bf B}\bdot\nabla\psi 
\equiv 0$). Note that we have added a toroidal magnetic field $B_{\varphi}\,\nabla\varphi$ in Eq.~\eqref{eq:B_axis}, with a covariant component 
$B_{\varphi}$ that is constant on a given magnetic-flux surface.

We first calculate the guiding-center toroidal canonical momentum from the guiding-center phase-space Lagrangian \eqref{eq:Gamma_gc_primitive}:
\begin{eqnarray}
P_{{\rm gc}\varphi} & \equiv & \left[ \frac{e}{\epsilon\,c}\;{\bf A} + p_{\|}\,\bhat - \epsilon\;J\,\left({\bf R} + \frac{1}{2}\,\nabla\btimes\bhat 
\right) \right]\bdot\pd{\bf X}{\varphi} \nonumber \\
 & = & -\frac{e\psi}{\epsilon\,c} + p_{\|}\,b_{\varphi} - \epsilon\,J \left[ b_{\sf z} + \bhat\bdot\nabla\btimes\left( \frac{1}{2}\,{\cal R}^{2}\,\nabla\varphi\right) \right] \nonumber \\
 &  &-\; \epsilon\,J \nabla\bdot\left(\bhat\btimes \frac{1}{2}\,{\cal R}^{2}\,\nabla\varphi\right)
\label{eq:Pgc_phi_def}
\end{eqnarray}
where we used ${\bf R}\bdot\partial{\bf X}/\partial\varphi \equiv b_{\sf z}$ \cite{RGL_1983} (i.e., the component of $\bhat$ along the symmetry axis $\wh{\sf z}$ for toroidal rotations), we wrote $\partial{\bf X}/\partial\varphi \equiv {\cal R}^{2}\,\nabla\varphi$ in terms of the major radius ${\cal R} \equiv |\nabla\varphi|^{-1}$, and we used the identity ${\bf F}\bdot\nabla\btimes{\bf G} \equiv \nabla\bdot({\bf G}\btimes{\bf F}) + {\bf G}\bdot\nabla\btimes{\bf F}$, for arbitrary vector fields ${\bf F}$ and ${\bf G}$. Next, we use
\[ \bhat\bdot\nabla\btimes\left( \frac{1}{2}\,{\cal R}^{2}\,\nabla\varphi\right) \;=\; \bhat\bdot\left(\wh{\cal R}\btimes\wh{\varphi}\right) \;=\; 
b_{\sf z}, \]
and
\[ \bhat\btimes \frac{1}{2}\,{\cal R}^{2}\,\nabla\varphi \;=\; \frac{1}{2B}\;\nabla\psi, \]
so that Eq.~\eqref{eq:Pgc_phi_def} becomes
\begin{eqnarray}
P_{{\rm gc}\varphi} & = & -\;\frac{e}{\epsilon\,c}\;\left[\psi \;+\; \epsilon^{2}\;\nabla\bdot\left(\frac{J}{2\,m\Omega}\,\nabla\psi\right) \right]
\nonumber \\
 &  &+\; p_{\|}\,b_{\varphi} \;-\; 2\,\epsilon\,J\;b_{\sf z}.
\label{eq:Pgc_phi}
\end{eqnarray}
Here, the second term on the first line in Eq.~\eqref{eq:Pgc_phi} is the second-order finite-Larmor-radius (FLR) correction to the first term. 

We now show that Eq.~\eqref{eq:Pgc_phi} is the exact guiding-center representation of the toroidal canonical angular momentum:
\begin{equation}
P_{{\rm gc}\varphi} \equiv {\sf T}_{\rm gc}^{-1}P_{\varphi} = -\,\frac{e}{\epsilon c}\,{\sf T}_{\rm gc}^{-1}\psi + {\sf T}_{\rm gc}^{-1}\left(m\,
{\bf v}\bdot\pd{\bf x}{\varphi} \right),
\label{eq:Pgc_phi_push}
\end{equation}
which guarantees the conservation of guiding-center toroidal canonical angular momentum
\begin{equation}
\frac{d_{\rm gc}P_{{\rm gc}\varphi}}{dt} \;=\; \frac{d_{\rm gc}}{dt}\left({\sf T}_{\rm gc}^{-1}\frac{}{}P_{\varphi}\right) \;=\; {\sf T}_{\rm gc}^{-1}
\left(\frac{dP_{\varphi}}{dt} \right) \;\equiv\; 0. 
\label{eq:Pgcphi_dot}
\end{equation}
First, we note that, while the term ${\sf T}_{\rm gc}^{-1}P_{\varphi}$ in Eq.~\eqref{eq:Pgc_phi_push} contains contributions that are gyroangle-independent and contributions that are explicitly gyroangle-dependent, the term $P_{{\rm gc}\varphi}$ is explicitly gyroangle-independent. Hence, the gyroangle-dependent contributions must vanish at all orders in 
$\epsilon$, and thus $P_{{\rm gc}\varphi} \equiv \langle{\sf T}_{\rm gc}^{-1}P_{\varphi}\rangle$; this identity, which is equivalent to a toroidal-canonical-momentum constraint on the guiding-center transformation, will be proved elsewhere \cite{Brizard_Tronko_2015}.

Secondly, we therefore introduce the guiding-center magnetic flux $\psi_{\rm gc} \equiv \langle {\sf T}_{\rm gc}^{-1}\psi\rangle$:
\begin{eqnarray}
\psi_{\rm gc}  & = & \psi \;+\; \epsilon^{2} \left( \langle\vb{\rho}_{1}\rangle\bdot\nabla\psi \;+\; \frac{1}{2}\,\langle\vb{\rho}_{0}\vb{\rho}_{0}\rangle:\nabla\nabla \psi \right) + \cdots \nonumber \\
 & = & \psi + \epsilon^{2}\;\nabla\bdot\left( \frac{J}{2\,m\Omega}\;\nabla\psi\right) + \epsilon^{2}\;\bhat\btimes\frac{{\bf v}_{\rm gc}}{\Omega}\bdot\nabla\psi,
\label{eq:psi_gc}
\end{eqnarray}
where we used Eqs.~\eqref{eq:rho1_ave}-\eqref{gc_pol_PK}. In Eq.~\eqref{eq:psi_gc}, the second term is an FLR correction to the first term, while the last term is easily recognized as a correction due to the guiding-center polarization \eqref{gc_pol_PK}. 

Thirdly, using the identity $\nabla\psi \equiv {\bf B}\btimes\partial{\bf X}/\partial\varphi$, with $\bhat\bdot{\bf v}_{\rm gc} \equiv 0$, we obtain a term proportional to the toroidal component of the guiding-center velocity:
\[ \bhat\btimes\frac{{\bf v}_{\rm gc}}{\Omega}\bdot\nabla\psi \;=\; \frac{B}{\Omega}\;\left({\bf v}_{\rm gc}\bdot\pd{\bf X}{\varphi}\right) \;\equiv\; 
\frac{B}{\Omega}\;v_{{\rm gc}\varphi}. \]
Hence, the final expression for the guiding-center toroidal canonical momentum defined by Eq.~\eqref{eq:Pgc_phi} is
\begin{equation}
P_{{\rm gc}\varphi} = -\frac{e \psi_{\rm gc}}{\epsilon\,c} + m \left( \frac{d_{0}{\bf X}}{dt} + \epsilon\;\frac{d_{1}{\bf X}}{dt}
\right)\bdot\pd{\bf X}{\varphi} - 2\,\epsilon\;J\,b_{\sf z},
\label{eq:Pgc_phi_final}
\end{equation}
where $d_{0}{\bf X}/dt \equiv (p_{\|}/m)\,\bhat$ and $d_{1}{\bf X}/dt \equiv {\bf v}_{\rm gc}$, while
\[ m\;\left( \frac{d_{0}{\bf X}}{dt} \;+\; \epsilon\;\frac{d_{1}{\bf X}}{dt}\right)\bdot\pd{\bf X}{\varphi} \;\equiv\; m\;{\cal R}^{2}\;
\frac{d_{\rm gc}\varphi}{dt} \]
denotes the guiding-center toroidal momentum with first-order corrections due to the guiding-center magnetic-drift velocity. 

The last term in Eq.~\eqref{eq:Pgc_phi_final} might be puzzling until we consider the guiding-center transformation of the particle toroidal canonical momentum $P_{{\rm gc}\varphi} \equiv \langle{\sf T}_{\rm gc}^{-1}\;p_{\varphi}\rangle$:
\begin{eqnarray}
P_{{\rm gc}\varphi} & = & -\,\frac{e}{\epsilon\,c}\;\langle{\sf T}_{\rm gc}^{-1}\psi\rangle \;+\; m\;\left\langle \left({\sf T}_{\rm gc}^{-1}
\frac{d{\bf x}}{dt}\right)\bdot\left({\sf T}_{\rm gc}^{-1}\pd{\bf x}{\varphi}\right)\right\rangle \nonumber \\
 & = & -\;\frac{e\psi_{\rm gc}}{\epsilon\,c} \;+\; m \left( \frac{d_{0}{\bf X}}{dt} \;+\; \epsilon\;\frac{d_{1}{\bf X}}{dt}
\right)\bdot\pd{\bf X}{\varphi} \nonumber \\
 &  &+\; \epsilon\;m\Omega\left\langle\pd{\vb{\rho}_{0}}{\theta}\bdot\pd{\vb{\rho}_{0}}{\varphi}\right\rangle + \cdots.
\label{eq:P_particle_phi}
\end{eqnarray}
Since $\partial\vb{\rho}_{0}/\partial\varphi \equiv \wh{\sf z}\btimes\vb{\rho}_{0}$ in axisymmetric magnetic geometry, the last term in 
Eq.~\eqref{eq:P_particle_phi} becomes
\[ \epsilon\;m\Omega\left\langle\pd{\vb{\rho}_{0}}{\theta}\bdot\pd{\vb{\rho}_{0}}{\varphi}\right\rangle \;=\; -\; 2\,\epsilon\;J\,b_{\sf z}, \]
and we recover the guiding-center toroidal canonical momentum \eqref{eq:Pgc_phi_final} from the guiding-center transformation of the particle toroidal canonical momentum \eqref{eq:P_particle_phi}.

\subsection{Comparison with Littlejohn's results}

By comparison, the guiding-center toroidal canonical momentum obtained by Littlejohn \cite{RGL_1983} and all subsequent guiding-center theories, is calculated with the choice 
$\vb{\Pi}_{1\bot} \equiv 0$:
\begin{equation}
(P_{{\rm gc}\varphi})_{\rm RGL} = -\;\frac{e\psi}{\epsilon\,c} + p_{\|}\,b_{\varphi} + \epsilon\,\left(\Pi_{1\|}\,b_{\varphi} \;-\frac{}{} J\,b_{\sf z}\right),
\label{eq:Pgc_phi_RGL}
\end{equation}
where the FLR correction to $\psi$ and the missing additional $b_{\sf z}$-term are hidden in $\Pi_{1\|}\,b_{\varphi} \equiv -\,\frac{1}{2}\,J\,\tau\,b_{\varphi}$:
\begin{eqnarray*} 
-\;\frac{1}{2}\,J\;\tau\,b_{\varphi} & = & -\,\frac{1}{2}\,J \left( \nabla\btimes\bhat\bdot\pd{{\bf X}}{\varphi} \;+\;
\bhat\btimes\pd{\bf X}{\varphi}\bdot\vb{\kappa} \right) \\
 & = & -\;\nabla\bdot\left(\frac{J}{2B}\;\nabla\psi\right) \;-\; J\;b_{\sf z} \;-\; \frac{J}{2B}\,\vb{\kappa}\bdot\nabla\psi.
\end{eqnarray*}
Hence, the Littlejohn guiding-center toroidal canonical momentum \eqref{eq:Pgc_phi_RGL} becomes
\begin{eqnarray}
(P_{{\rm gc}\varphi})_{\rm RGL} & = & -\;\frac{e}{\epsilon\,c}\;\left[\psi \;+\; \epsilon^{2}\;\nabla\bdot\left(\frac{J}{2\,m\Omega}\,\nabla\psi\right)  \right] \nonumber \\
 &  &+\; p_{\|}\,b_{\varphi} - \epsilon\left(2\,J\;b_{\sf z} + \frac{J\vb{\kappa}}{2B}\bdot\nabla\psi\right).
\label{eq:Pgc_phi_RGL_FLR}
\end{eqnarray}

The Littlejohn guiding-center toroidal canonical momentum \eqref{eq:Pgc_phi_RGL_FLR}, of course, has the same form as Eq.~\eqref{eq:Pgc_phi_final} since its associated guiding-center magnetic flux is
\begin{eqnarray} 
(\psi_{\rm gc})_{\rm RGL} & = & \psi \;+\; \epsilon^{2}\;\nabla\bdot\left( \frac{J}{2\,m\Omega}\;\nabla\psi\right) \nonumber \\
 &  &+\; \epsilon^{2}\;\left(\bhat\btimes\frac{{\bf v}_{\rm gc}}{\Omega} \;+\; \frac{J\vb{\kappa}}{2m\Omega}\right)\bdot\nabla\psi \nonumber \\
 & \equiv & \psi_{\rm gc} \;+\; \epsilon^{2}\;\frac{J\,\vb{\kappa}}{2\,m\Omega}\bdot\nabla\psi,
\end{eqnarray}
where the extra term associated with the normal magnetic curvature $\vb{\kappa}\bdot\nabla\psi/|\nabla\psi|$ was eliminated by our choice 
\eqref{eq:Pi1_perp_choice} for $\vb{\Pi}_{1\bot}$. Once again, we point out that this choice is mandated by a correct derivation of the standard guiding-center polarization \eqref{gc_pol_PK}. 

We note that Belova {\it et al.} \cite{Belova} have shown that the second-order $(\epsilon^{2})$ corrections to the guiding-center toroidal canonical momentum \eqref{eq:Pgc_phi_RGL} were shown to be crucial in obtaining excellent conservation properties of toroidal canonical momentum in realistic axisymmetric tokamak plasmas.
It remains to be seen whether the additional guiding-center polarization correction \eqref{eq:Pi1_perp_choice} yields improved comservation properties.

\section{\label{sec:Sum}Summary}

In conclusion, a systematic derivation of the Hamiltonian guiding-center dynamics has been derived by Lie-transform perturbation analysis. The guiding-center Poisson bracket derived from the guiding-center phase-space Lagrangian \eqref{eq:Gamma_gc_primitive} and the guiding-center Hamiltonian \eqref{eq:Ham_gc_final}. These guiding-center Hamilton equations have passed several consistency tests along the way. 

First, we verified that our guiding-center transformation satisfies the guiding-center Jacobian constraints at first and second orders. Next, we verified that our guiding-center transformation also satisfies the guiding-center Lagrangian constraints at first and second orders. In fact, the use of the Lagrangian constraints on the guiding-center transformation yields a natural expression \eqref{eq:Ham_gc_final} for the guiding-center Hamiltonian in terms of the guiding-center velocity $d_{\rm gc}{\bf X}/dt$ and the guiding-center displacement velocity $d_{\rm gc}\vb{\rho}_{\rm gc}/dt$. When the polarization term $\vb{\Pi}_{1\bot}$ is ignored in the guiding-center Hamiltonian, our second-order guiding-center Hamiltonian is identical to the Hamiltonian derived by Burby, Squire, and Qin \cite{Burby_SQ_2013}.

We also showed that the perpendicular component of $\vb{\Pi}_{1}$, which could not be determined at the perturbation orders considered in this work, could nevertheless not be chosen to be zero, in contrast to the simplifying choice made by Littlejohn \cite{RGL_1983}. The choice \eqref{eq:Pi1_final} defined in the present work not only yields the standard Pfirsch-Kaufman guiding-center polarization \eqref{gc_pol_PK}, but also yields a simpler and more transparent guiding-center representation of the particle toroidal canonical momentum \eqref{eq:Pgc_phi_final}. 

Work by AJB was partially supported by a U.~S.~DoE grant under contract No.~DE-SC0006721. This work has been carried out within the framework of the EUROfusion Consortium and has received funding from the Euratom research and training programme 2014-2018 under grant agreement No 633053. The views and opinions expressed herein do not necessarily reflect those of the European Commission.


\begin{thebibliography}{99}
\bibitem{footnote_2} We note, however, that the second-order guiding-center Hamiltonian $\Psi_{2(PC)}$ does not satisfy the second-order Hamiltonian constraint \eqref{eq:Hamiltonian_constraint_2}, since standard Lie-transform perturbation theory [e.g., see Eq.~\eqref{eq:sigma3_def}] explicitly assumes that the Lagrangian gauge-invariance property $\Gamma_{\rm gc}^{\prime} \equiv \Gamma_{\rm gc} + \exd\sigma$ involves gyroangle-dependent gauge functions only (i.e., $\langle\sigma\rangle \equiv 0$).
\end{thebibliography}
\end{document}